\documentclass[superscriptaddress,showkeys,showpacs,
nofootinbib,byrevtex,preprintnumbers]{revtex4}  
\usepackage{graphicx}
\begin{document}      
\preprint{YITP-07-96}
\preprint{PNU-NTG-11/2007}
\title{Pion electromagnetic form factor at finite density
\footnote{This report is prepared for the proceeding of the talk
  "Effective QCD via instanton", given at the workshop {\it
    Strangeness in multi-quark systems} which was held during 26--28
  November, 2007 in Sendai, Japan.}} 
%-------------------------------------------------
\author{Seung-il Nam}
\email[E-mail: ]{sinam@yukawa.kyoto-u.ac.jp}
\affiliation{Yukawa Institute for Theoretical Physics (YITP), Kyoto
University, Kyoto 606-8502, Japan} 
%-------------------------------------------------
\author{Hyun-Chul Kim}
\email[E-mail: ]{hchkim@pusan.ac.kr}
\affiliation{Department of Physics and Nuclear physics \& Radiation
Technology Institute (NuRI), Pusan National University, Busan
609-735, Korea}  
%-------------------------------------------------
\date{\today}
%-------------------------------------------------
\begin{abstract}
In this talk, we introduce an effective chiral action, derived from
the instanton QCD vacuum configuration, in order to study the
modification of mesonic properties in medium.  We present our recent
theoretical results on the electromagnetic from factor and
$\rho$-meson mass shift at finite density.  We observe that the
$\rho$-meson mass is dropped by about $10\%$: 
$m^*_{\rho}/m_{\rho}\approx1-0.12\,\mu_B/\mu_0$, which is compatible 
with that estimated by the QCD sum rule as well as by Brown and Rho. 
\end{abstract} 
%-------------------------------------------------
\maketitle
%-------------------------------------------------
%-------------------------------------------------
\section{Introduction}
%-------------------------------------------------
It is worth mentioning that the nonlocal chiral quark model
(N$\chi$QM), derived from the instanton QCD vacuum, has
been successfully applied to describing low-energy properties of hadrons
~\cite{Diakonov:1985eg,Schafer:1993ra,Diakonov:2002fq,Musakhanov:1998wp,
Kim:2004hd,Goeke:2007nc}.  Especially for the pseudoscalar (PS)
mesonic sector, we have carried out several theoretical works which
turn out to describe experimental data quantitatively well and are
compatible with other model
calculations~\cite{Nam:2006au,Nam:2006mb,Nam:2006sx,Nam:2007fx, 
Nam:2007gf}.  The effective chiral action consists of quark and
(anti)instanton interaction, encoded in the t'Hooft 
$2N_f$-interaction~\cite{Diakonov:2002fq}. Having bosonized
it, the effective chiral action is reconstructed in terms of the quark
and nonlinear PS meson background fields.  We start directly 
with the Euclidean effective chiral action in the presence of the
external electromagnetic field, which was constructed to satisfy the
gauge invariance, explicitly as 
follows~\cite{Musakhanov:2002xa,Kim:2004hd,Goeke:2007nc}:   
%EQUATION>>>
\begin{equation}
\label{eq:ECA}
{\cal S}_{\rm eff}[\pi,\hat{m},V]=
-{\rm Sp}\ln\left[i\rlap{/}{D}+i\hat{m}+\sqrt{M(iD)}U_5\sqrt{M(iD)} \right],
\end{equation}
%EQUATION<<<
where the ${\rm Sp}$ stands for the functional trace running over
space-time $\int d^4 x \langle x|\cdots|x\rangle$, color, flavor and
Dirac spin spaces. The $D$ denotes the covariant derivative with an
external vector current: $iD_{\mu}=i\partial_{\mu}+V_{\mu}$.  The
$\hat{m}$ is the current-quark mass matrix, ${\rm diag}(m_u,m_d)$
considering isospin symmetry ($m_u=m_d$).  The nonlinear background
Goldstone boson field, $U_5$ is defined as:   
%EQUATION>>>
\begin{equation}
U_5=U(x)\frac{1+\gamma _{5}}{2}+U^{\dagger }(x)
\frac{1-\gamma_{5}}{2}
=1+\frac{i}{F_{\pi}}\gamma _{5}\pi\cdot
\lambda-\frac{1}{2F_{\pi}^{2}}(\pi\cdot\lambda)^2\cdots,
\end{equation} 
%EQUATION<<<
where $F_{\pi}$ stands for the pion weak decay constant, set to be
$93$ MeV. The momentum-dependent quark mass, $M(k)$ arises from 
the Fourier transform of the quark zero-mode solution $\Psi_0$: 
%EQUATION>>>
\begin{equation}
\label{eq:QZMS}
\left[i\rlap{/}{\partial}+\rlap{/}{A}_{I\bar{I}}\right]\Psi_0=0,
\end{equation}
%EQUATION<<< 
where the $A^{\mu}_{I\bar{I}}$ is the (anti)instanton solution
satisfying the self-dual condition for the gluon field strength
tensor,
$G_{\mu\nu}(x)_{I\bar{I}}=\pm\tilde{G}_{\mu\nu}(x)_{I\bar{I}}$. Throughout
the present talk, we exploit the singular-gauge solution  for 
$A^{\mu}_{I\bar{I}}$~\cite{Diakonov:2002fq}. For simplicity and better
numerical calculation, we introduce a parameterization of the
dipole-type form factor for the momentum-dependent quark mass: 
%EQUATION>>>
\begin{equation}
\label{eq:MDQM}
M(k)=M_0\left[\frac{2\Lambda^2}{k^2+2\Lambda^2}\right]^2,
\end{equation}
%EQUATION<<<
where $\Lambda\approx1/\bar{\rho}\approx600$ MeV, and $M_0$ is
determined self-consistently via the saddle-point equation of the
model, resulting in about $350$ MeV as usual for the mesonic
sector.  We want to emphasize on the fact that there is no adjustable 
free parameters within the model. In other words, the  present model
is highly constrained.  
%-------------------------------------------------
\section{Pion electromagnetic form factor at finite density}
\begin{figure}[t]
\begin{tabular}{cc}
\includegraphics[width=7.5cm]{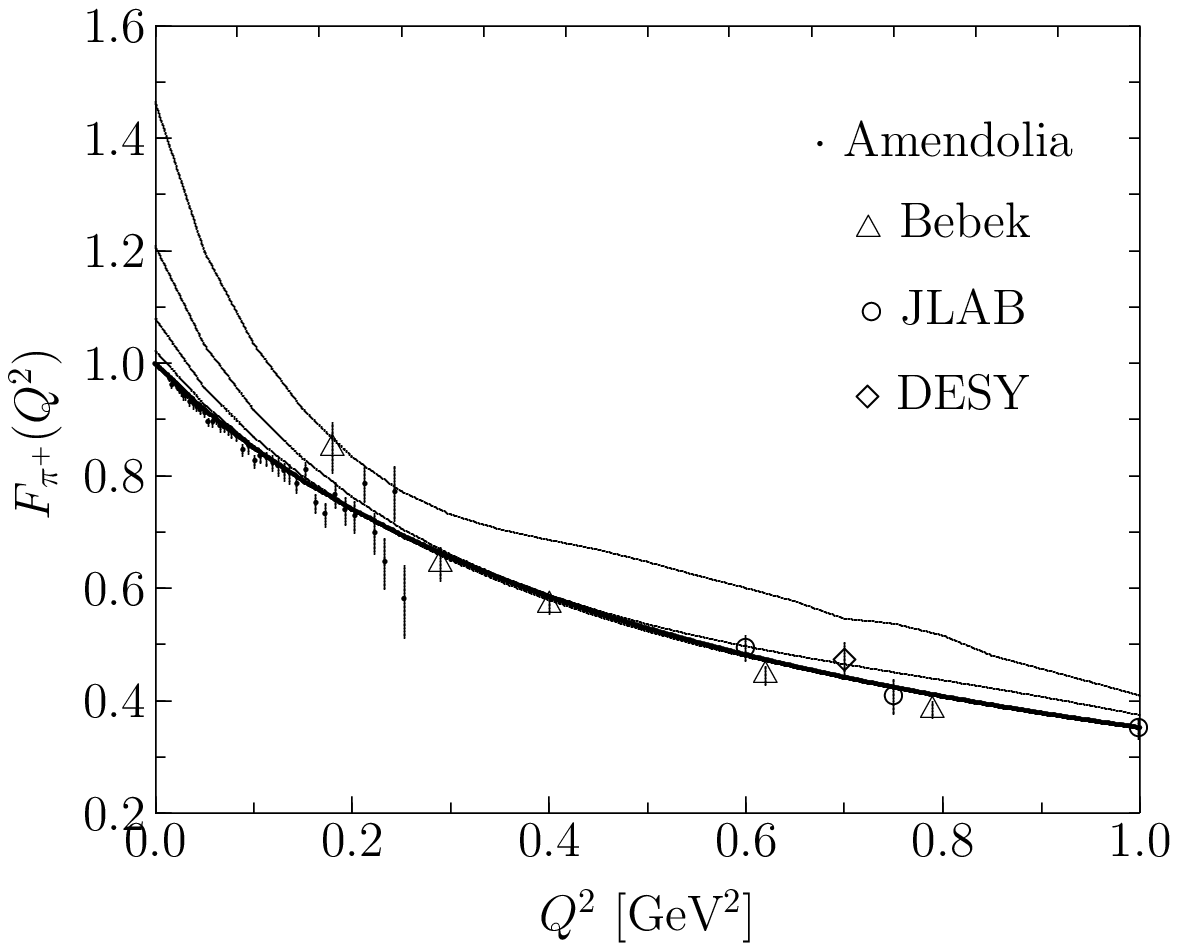}
\includegraphics[width=7.5cm]{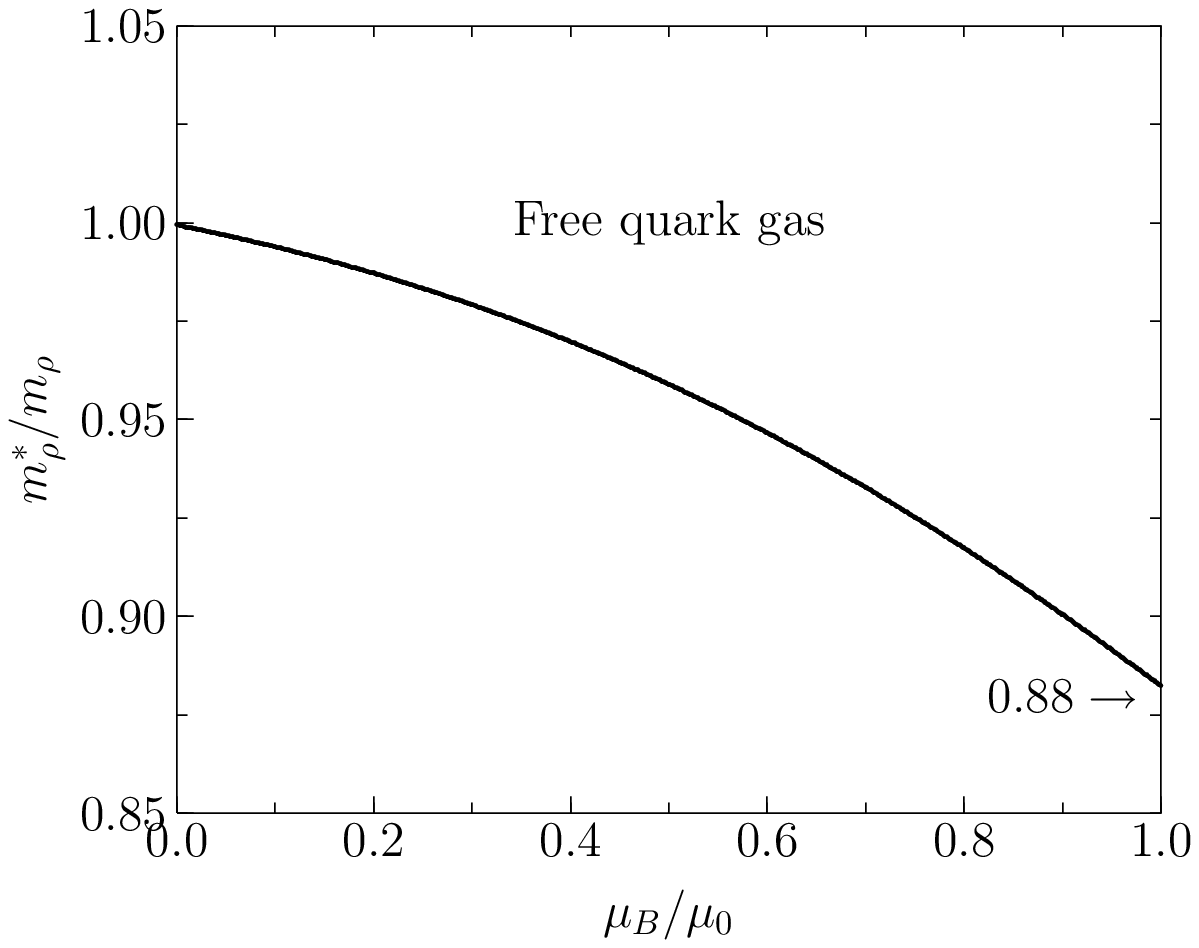}
\end{tabular}
\caption{Left panel: pion electromagneti  form factor with respect to
the quark chemical potential, $\mu_q$. Experimental data are taken
from
Refs.~\cite{Bebek:1974iz,Bebek:1974ww,Amendolia:1986wj,Volmer:2000ek}. Right
panel: the $\rho$-meson mass ratio with respect to the ratio
$\mu_B/\mu_0$.} 
\label{fig0}
\end{figure}

%-------------------------------------------------
In this section, we briefly explain how to compute the pion EM FF at
nonzero quark density as in Ref.~\cite{Nam:2007gf}.  In
the left panel of Figure~\ref{fig0} we depict the numerical result for
the EM FF for the vacuum ($\mu_q=0$, thick solid line), and it is well
reproduced in comparison to available experimental data.  Moreover,
the radius $\langle r^2\rangle$ calculated from the EM FF becomes
$0.454\,{\rm fm}^2$, which is in good agreement with the experimental
value $\sim0.445\,{\rm fm}^2$~\cite{Yao:2006px}.  

Now, we are in a position to discuss the way to include the quark
number chemical potential ($\mu_q$) in the model. Since we are in the
chiral limit for $N_f=2$, the isovector part of the quark number
chemical potential is zero. Hence, we will consider only the isoscalar
one hereafter. As pointed out in Refs.~\cite{Carter:1998ji}, a simple
replacement is enough for this purpose: $i\partial\to i\partial
-i\mu$, where $\mu=(\vec{0},\mu_q)$. Then the effective chiral action
of Eq.~(\ref{eq:ECA}) can be modified as follows: 
%EQUATION>>>
\begin{eqnarray} 
\label{eq:ECA1} 
{\cal S}_{\rm eff}[\pi,\mu, V]
=-{\rm Sp}\ln\left[i\rlap{/}{D}'
+i\sqrt{{\cal M}(iD')} U_5\sqrt{{\cal M}(iD')}\right],
\end{eqnarray} 
%EQUATION<<<
where $iD'_{\mu}=iD_{\mu}-i\mu_q$. The momentum-dependent quark mass
is also modified with the nonzero $\mu_q$ from Eq.~(\ref{eq:MDQM}) as
follows: 
%EQUATION>>>
\begin{eqnarray}
\label{eq:MFD}
{\cal M}(i\partial,\mu)&=&{\cal M}_0
\left[\frac{2\Lambda^2}{(i\partial-i\mu)(i\partial-i\mu)+2\Lambda^2}
\right]^2={\cal M}_0{\cal F}^2(i\partial,\mu).
\end{eqnarray}
%EQUATION<<<
Here we assume the small density limit, we take ${\cal M}_0$ to remain
intact in medium.  We verified that this simple form factor
reproduce the exact zero mode with the finite $\mu_q$ qualitatively
well. With this modification of the effective chiral action, 
we obtain the numerical  results for the pion EM FF with respect to
the nonzero $\mu_q$ (thin solid lines) as depicted in the left panel
of Figure~\ref{fig0}.  We draw them for the $\mu_q=50$, $100$, $150$
and $200$ MeV separately. We observe that the value of the pion EM FF
at $Q^2=0$ becomes larger as the $\mu_q$ 
increases. At the same time, the slope at $Q^2=0$, which
corresponds to the pion EM charge radius, increases with respect to
the $\mu_q$.  
%FIGURE>>>
%FIGURE<<<

%-------------------------------------------------
\section{$\rho$-meson mass shift in medium}
%-------------------------------------------------
The EM FF of the pion in the time-like region ($Q^2>0$) can be written
in a simple parameterization in terms of the vector meson dominance
(VMD) as follows~\cite{Sakurai}: 
%EQUATION>>>
\begin{eqnarray}
\label{eq:EMFFVMD}
F(Q^2)\approx\frac{{\cal C}\,m^2_{\rho}}
{m^2_{\rho}-Q^2+i\Gamma_{\rho}m_{\rho}},
\end{eqnarray}
%EQUATION<<<
where $m_{\rho}$ and $\Gamma_{\rho}$ denote the mass and full decay
width of the $\rho$ meson in free space.  The coefficient ${\cal C}$ 
is introduced for the normalization to match with the experimental
data. For instance, these values can be chosen as $m_{\rho}=0.77$ GeV,
$\Gamma_{\rho}=0.135$ GeV and ${\cal C}=1.3$ to produce the dilepton
data~\cite{Schulze:2003za} in free space.  Now, we make an 
assumption to proceed the discussion: Even in the case of the nonzero
quark number chemical potential ($\mu_q\ne0$), the EM FF of the pion
follows the same form of Eq.~(\ref{eq:EMFFVMD}) by only replacing
$m_{\rho}\to m^*_{\rho}$, $\Gamma_{\rho}\to \Gamma^*_{\rho}$ and
${\cal C}\to {\cal C}^*$.  Note that the superscripted asterisk ($*$)
indicates that they depend on the $\mu_q$.  More generally, they must
be expressed as fucntions of $\mu_q$ as well as of the temperature
$T$. The situation considered in the present work is, however, chosen
to be $T=0$ for simplicity hereafter. Then, the $\rho$-meson mass for
the nonzero $\mu_q$ can be obtained by the following expression:  
%EQUATION>>>
\begin{eqnarray}
\label{eq:mrho}
m^*_{\rho}=m_{\rho}\left[\frac{0.45\,{\rm fm}^2\times{\cal C}^*}
{\langle r^2\rangle^*}\right]^{1/2},
\end{eqnarray}
%EQUATION<<<
where $\langle r^2\rangle^*$ and ${\cal C}^*$ can be computed using
the N$\chi$QM as shown previously. In the right panel of
Figure~\ref{fig0} we show the corresponding numerical results, where
$\mu_B$ denotes the baryon number chemical potential being equal to 
$3\mu_q$.  The $\mu_0$ corresponds to the normal nuclear density 
$\rho_0=0.17\,{\rm fm}^{-3}$, and is chosen to be about $300$ MeV,
the system being assumed in free quark-gas phase.  We clearly see that
the $\rho$-meson mass decreases with respect to the quark chemical 
potential, $\mu_q$. To see this behavior in a qualitative manner, we
employ the following linear parameterization for the mass shift of the
$\rho$-meson: 
%EQUATION>>>
\begin{eqnarray}
\label{eq:para}
\frac{m^*_{\rho}}{m_{\rho}}\approx1-\alpha\,\frac{\mu_B}{\mu_0},
\end{eqnarray}
%EQUATION<<<
 If we determine the value of $\alpha$ from the end points
 ($\mu_B/\mu_0=0,1$), we have $\alpha=0.12$. In Ref.~\cite{Hatsuda:1991ez},
 $\alpha$ was estimated to be around $0.15\sim0.18$, which is similar
 to that based on phenomenological method in 
 Ref.~\cite{Brown:1991kk,Brown:1991wb}.  In contrast, the $\rho$-meson
 mass drops suddenly as the $\mu_q$ increases in
 Ref.~\cite{Muroya:2002ry} using the lattice simulation in terms of
 SU(2) color symmetry.  It was also noted that, from the chiral
 models, the $\rho$-meson pole moves very differently in the complex
 plane for $F^*_{\pi}\approx F_{\pi}$~\cite{Yokokawa:2002pw}.  
%-------------------------------------------------
\section{Summary and conlcusion}
%-------------------------------------------------
In this talk, we have investigated the $\rho$-meson mass shift in
medium in terms of the pion electromagnetic charge radius, which was
computed as a function of the quark number chemical potential within
the framework of the nonlocal chiral quark model. In addition, the
vector meson dominance was employed to express the pion EM charge
radius associated with the $\rho$-meson mass. When we took a
phenomenological parameterization for the mass shift, 
$m^*_{\rho}/m_{\rho}=1-\alpha\,\mu_B/\mu_0$, $\alpha$ turned out to be
about $0.12$ which was compatible to the estimate from the QCD sum
rule and phenomenological  estimations ($0.15\sim0.18$). Although we
note that there are ambiguities in theoretical calculations, they will
not alter the qualitative consequences much. A detailed work on the 
present talk will appear elsewhere~\cite{Nam:2008xx}. 
%-------------------------------------------------
\section*{Acknowledgments}
%-------------------------------------------------
S.i.N. is grateful to the organizers for the workshop {\it Strangeness
  in multi-quark systems}, which was held during 26--28 November, 2007
in Sendai, Japan (http://nexus.kek.jp/Tokutei/workshop/2007). Authors
are also thankful to C.~H.~Lee, S.~H.~Lee, M.~Oka, T.~Kunihiro and
Y.Kwon for fruitful discussions. The work of S.i.N. is partially
supported by the grant for Scientific Research (Priority Area
No. 17070002) from the Ministry of Education, Culture, Science and
Technology, Japan. The work of H.Ch.K. is supported by the Korea
Research Foundation Grant funded by the Korean Government(MOEHRD)
(KRF-2006-312-C00507).     
%--------------------------------------------------

%--------------------------------------------------
\end{document}